\documentclass{aastex6}

\fullcollaborationName{The SEEDS team}

\begin{document}

\title{A Resolved Near-Infrared Image of The Inner Cavity in The GM Aur Transitional Disk\footnote{Based on IRCS and HiCIAO data collected at Subaru Telescope, operated by the National Astronomical Observatory of Japan.}}

\author{Daehyeon OH\altaffilmark{1,2,3}}
\affil{SOKENDAI (Department of Astronomical Science, The Graduate University for Advanced Studies) \\ 2-21-1 Osawa, Mitaka, Tokyo, 181-8588, Japan}

\author{Jun HASHIMOTO}
\affil{Astrobiology Center of NINS\\ 2-21-1, Osawa, Mitaka, Tokyo, 181-8588, Japan}

\author{Joseph C. CARSON}
\affil{Department of Physics and Astronomy, College of Charleston \\ 66 George St., Charleston, SC 29424, USA}

\author{Markus JANSON}
\affil{Department of Astronomy, Stockholm University, AlbaNova University Center \\ SE-106 91 Stockholm, Sweden}

\author{Jungmi KWON}
\affil{Institute of Space and Astronautical Science, Japan Aerospace Exploration Agency \\ 3-1-1 Yoshinodai, Chuo, Sagamihara, Kanagawa 252-5210, Japan}

\author{Takao NAKAGAWA}
\affil{Institute of Space and Astronautical Science, Japan Aerospace Exploration Agency \\3-1-1 Yoshinodai, Chuo, Sagamihara, Kanagawa 252-5210, Japan}

\author{Satoshi MAYAMA}
\affil{The Center for the Promotion of Integrated Sciences, SOKENDAI (Department of Astronomical Science, The Graduate University for Advanced Studies), Shonan International Village, Hayama-cho, Miura-gun, Kanagawa 240-0193, Japan}

\author{Taichi Uyama}
\affil{Department of Astronomy, The University of Tokyo \\ 7-3-1, Hongo, Bunkyo-ku, Tokyo, 113-0033, Japan}

\author{Yi Yang}
\affil{Department of Astronomical Science, The Graduate University for Advanced Studies (SOKENDAI) \\ 2-21-1 Osawa, Mitaka, Tokyo, 181-8588, Japan}

\author{Tomoyuki KUDO}
\affil{Subaru Telescope, National Astronomical Observatory of Japan \\ 650 North A'ohoku Place, Hilo, HI96720, USA}

\author{Nobuhiko KUSAKABE}
\affil{Astrobiology Center of NINS\\ 2-21-1, Osawa, Mitaka, Tokyo, 181-8588, Japan}

\author{Lyu ABE}
\affil{Laboratoire Lagrange (UMR 7293), Universite de Nice-Sophia Antipolis, CNRS, Observatoire de la Coted'azur \\ 28 avenue Valrose, 06108 Nice Cedex 2, France}

\author{Eiji AKIYAMA}
\affil{National Astronomical Observatory of Japan \\ 2-21-1, Osawa, Mitaka, Tokyo, 181-8588, Japan}

\author{Wolfgang BRANDNER}
\affil{Max Planck Institute for Astronomy, \\ K\"{o}onigstuhl 17, 69117 Heidelberg, Germany}

\author{Timothy D. BRANDT}
\affil{Astrophysics Department, Institute for Advanced Study \\ Princeton, NJ, USA}

\author{Thayne CURRIE}
\affil{Subaru Telescope, National Astronomical Observatory of Japan \\ 650 North A'ohoku Place, Hilo, HI96720, USA}

\author{Markus FELDT}
\affil{Astrophysics Department, Institute for Advanced Study \\ Princeton, NJ, USA}

\author{Miwa GOTO}
\affil{Universitats-Sternwarte Munchen, Ludwig-Maximilians-Universitat, \\ Scheinerstr. 1, D-81679 Munchen,Germany}

\author{Carol A. GRADY\altaffilmark{4,5}}
\affil{Exoplanets and Stellar Astrophysics Laboratory, Code 667, Goddard Space Flight Center \\  Greenbelt, MD 20771, USA}

\author{Olivier GUYON}
\affil{Subaru Telescope, National Astronomical Observatory of Japan \\ 650 North A'ohoku Place, Hilo, HI96720, USA}

\author{Yutaka HAYANO}
\affil{Subaru Telescope, National Astronomical Observatory of Japan \\ 650 North A'ohoku Place, Hilo, HI96720, USA}

\author{Masahiko HAYASHI}
\affil{National Astronomical Observatory of Japan \\ 2-21-1, Osawa, Mitaka, Tokyo, 181-8588, Japan}

\author{Saeko S. HAYASHI}
\affil{Subaru Telescope, National Astronomical Observatory of Japan \\ 650 North A'ohoku Place, Hilo, HI96720, USA}

\author{Thomas HENNING}
\affil{Max Planck Institute for Astronomy, \\ K\"{o}onigstuhl 17, 69117 Heidelberg, Germany}

\author{Klaus W. HODAPP}
\affil{Institute for Astronomy, University of Hawaii \\ 640 N. A'ohoku Place, Hilo, HI 96720, USA}

\author{Miki ISHII}
\affil{National Astronomical Observatory of Japan \\ 2-21-1, Osawa, Mitaka, Tokyo, 181-8588, Japan}

\author{Masanori IYE}
\affil{National Astronomical Observatory of Japan \\ 2-21-1, Osawa, Mitaka, Tokyo, 181-8588, Japan}

\author{Ryo KANDORI}
\affil{National Astronomical Observatory of Japan \\ 2-21-1, Osawa, Mitaka, Tokyo, 181-8588, Japan}

\author{Gillian R. KNAPP}
\affil{Department of Astrophysical Science, Princeton University \\ Peyton Hall, Ivy Lane, Princeton, NJ08544, USA}

\author{Masayuki KUZUHARA}
\affil{Department of Earth and Planetary Sciences, Tokyo Institute of Technology \\ 2-12-1 Ookayama, Meguro-ku, Tokyo 152-8551, Japan}

\author{Taro MATSUO}
\affil{Department of Astronomy, Kyoto University, Kitashirakawa-Oiwake-cho, Sakyo-ku, Kyoto, Kyoto 606-8502, Japan}

\author{Michael W. MCELWAIN}
\affil{Exoplanets and Stellar Astrophysics Laboratory, Code 667, Goddard Space Flight Center \\ Greenbelt, MD 20771, USA}

\author{Shoken MIYAMA}
\affil{Hiroshima University \\ 1-3-2, Kagamiyama, Higashihiroshima, Hiroshima 739-8511, Japan}

\author{Jun-Ichi MORINO}
\affil{National Astronomical Observatory of Japan \\ 2-21-1, Osawa, Mitaka, Tokyo, 181-8588, Japan}

\author{Amaya MORO-MARTIN\altaffilmark{6}}
\affil{Space Telescope Science Institute \\ 3700 San Martin Drive, Baltimore, MD 21218, USA}

\author{Tetsuo NISHIMURA}
\affil{Subaru Telescope, National Astronomical Observatory of Japan \\ 650 North A'ohoku Place, Hilo, HI96720, USA}

\author{Tae-Soo PYO}
\affil{Subaru Telescope, National Astronomical Observatory of Japan \\ 650 North A'ohoku Place, Hilo, HI96720, USA}

\author{Eugene SERABYN}
\affil{Kavli Institute for Physics and Mathematics of the Universe, The University of Tokyo \\ 5-1-5, Kashiwanoha, Kashiwa, Chiba 277-8568, Japan}

\author{Takuya SUENAGA\altaffilmark{1}}
\affil{Department of Astronomical Science, The Graduate University for Advanced Studies (SOKENDAI) \\ 2-21-1 Osawa, Mitaka, Tokyo, 181-8588, Japan}

\author{Hiroshi SUTO\altaffilmark{4}}
\affil{National Astronomical Observatory of Japan \\ 2-21-1, Osawa, Mitaka, Tokyo, 181-8588, Japan}

\author{Ryuji SUZUKI}
\affil{National Astronomical Observatory of Japan \\ 2-21-1, Osawa, Mitaka, Tokyo, 181-8588, Japan}

\author{Yasuhiro H. TAKAHASHI\altaffilmark{8}}
\affil{National Astronomical Observatory of Japan \\ 2-21-1, Osawa, Mitaka, Tokyo, 181-8588, Japan}

\author{Naruhisa TAKATO}
\affil{Subaru Telescope, National Astronomical Observatory of Japan \\ 650 North A'ohoku Place, Hilo, HI96720, USA}

\author{Hiroshi TERADA}
\affil{National Astronomical Observatory of Japan \\ 2-21-1, Osawa, Mitaka, Tokyo, 181-8588, Japan}

\author{Christian THALMANN}
\affil{Swiss Federal Institute of Technology (ETH Zurich), Institute for Astronomy \\ Wolfgang-Pauli-Strasse 27, CH-8093 Zurich, Switzerland}

\author{Edwin L. TURNER\altaffilmark{9}}
\affil{Department of Astrophysical Science, Princeton University \\ Peyton Hall, Ivy Lane, Princeton, NJ08544, USA}

\author{Makoto WATANABE}
\affil{Department of Cosmosciences, Hokkaido University \\ Kita-ku, Sapporo, Hokkaido 060-0810, Japan}

\author{Toru YAMADA}
\affil{Astronomical Institute, Tohoku University \\ Aoba-ku, Sendai, Miyagi 980-8578, Japan}

\author{Hideki TAKAMI}
\affil{National Astronomical Observatory of Japan \\ 2-21-1, Osawa, Mitaka, Tokyo, 181-8588, Japan}

\author{Tomonori USUDA}
\affil{National Astronomical Observatory of Japan \\ 2-21-1, Osawa, Mitaka, Tokyo, 181-8588, Japan}

\and

\author{Motohide TAMURA\altaffilmark{1,7}}
\affil{Department of Astronomy, The University of Tokyo \\ 7-3-1, Hongo, Bunkyo-ku, Tokyo, 113-0033, Japan}

\altaffiltext{1}{National Astronomical Observatory of Japan, 2-21-1, Osawa, Mitaka, Tokyo, 181-8588, Japan}
\altaffiltext{2}{National Meteorological Satellite Center, 64-18 Guam-gil, Gawnghyewon, Jincheon, Chungbuk, 27803, Korea}
\altaffiltext{3}{daehyun.oh@nao.ac.jp}
\altaffiltext{4}{Eureka Scientific, 2452 Delmer, Suite 100, Oakland CA96002, USA}
\altaffiltext{5}{Goddard Center for Astrobiology}
\altaffiltext{6}{Center for Astrophysical Sciences, Johns Hopkins University, Baltimore MD 21218, USA}
\altaffiltext{7}{Astrobiology Center of NINS, 2-21-1, Osawa, Mitaka, Tokyo, 181-8588, Japan}
\altaffiltext{8}{Department of Astronomy, The University of Tokyo, 7-3-1, Hongo, Bunkyo-ku, Tokyo, 113-0033, Japan}
\altaffiltext{9}{Kavli Institute for Physics and Mathematics of the Universe, The University of Tokyo, 5-1-5, Kashiwanoha, Kashiwa, Chiba 277-8568, Japan}

\begin{abstract}

  We present high-contrast $H$-band polarized intensity (PI) images of the transitional disk around the young solar-like star GM Aur. The near-infrared direct imaging of the disk was derived by polarimetric differential imaging using the Subaru 8.2 m Telescope and HiCIAO. An angular resolution and an inner working angle of $0\farcs07$ and $r\sim0\farcs05$, respectively, were obtained. We clearly resolved a large inner cavity, with a measured radius of 18$\pm$2 au, which is smaller than that of a submillimeter interferometric image (28 au). This discrepancy in the cavity radii at near-infrared and submillimeter wavelengths may be caused by a 3--4$M_{\rm Jup}$ planet about 20 au away from the star, near the edge of the cavity. The presence of a near-infrared inner cavity is a strong constraint on hypotheses for inner cavity formation in a transitional disk. A dust filtration mechanism has been proposed to explain the large cavity in the submillimeter image, but our results suggest that this mechanism must be combined with an additional process. We found that the PI slope of the outer disk is significantly different from the intensity slope obtained from HST/NICMOS, and this difference may indicate the grain growth process in the disk.

\end{abstract}


\section{Introduction}

The presence of circumstellar disks was predicted as an inevitable consequence of angular momentum conservation during gravitational collapse in star formation, and circumstellar disks  are now considered to be the initial phase of planetary evolution. Typical disks exhibit continuous infrared excess above the stellar emission, but some disks have deficits of near- and mid-infrared excesses and show nearly photospheric emission, while having substantial excesses at the far-infrared and beyond \citep{str89}. Spectral energy distribution (SED) modeling studies by the $Spitzer$ Infrared Spectrograph \citep[IRS;][]{hou04} and the $Infrared Space Observations$ \citep[ISO;][]{kes96} suggest that a spectral dent in the near infrared can be explained as being due to optically thin inner cavities with radii of 10 au or more in optically thick disks \citep{cal05,dal05}. Such disks are referred to as transitional disks, and have become central to understanding the evolutionary transition between protoplanetary and debris disks. The inside-out disk clearing in transitional disks must occur over a very short time scale ($<$ 10 Myr), since transitional disks only appear around young stellar objects with ages of 1--10 Myr. Also, the small frequency of transitional disks may reflect the fact that not all disks experience inner disk clearing at young ages \citep{muz10}. However, it is not well understood which physical mechanisms drive the disk clearing process.

One promising clearing mechanism is dust filtration by outer edge of a planet-induced gap \citep{ric06,zhu12}. This mechanism was proposed to explain the dust dynamics and gas dynamics in the disk. Since the radial density and the pressure gradient of the gas are negative at the outer edge of a planet-induced gap, dust particles drift outward by gaining angular momentum from the gas. When drifting dust particles overcome the coupling with the gas, they stay at the outer edge of the gap while the gas falls into the star through the gap. This process acts like a dust filter, and it depletes the inner dust disk to the radius of the planet-induced gap, producing a dustless inner cavity. It should be noted that this mechanism is effective only if the dust particles are large enough to have significant drift velocities. \citet{zhu12} showed that particles larger than 0.01--0.1 mm can be filtered by a \deleted{reasonable} gap. \citet{don12a} also suggested that the dust filtration effect is a promising mechanism for explaining why there is no sign of cavities in their near-infrared scattered light images for many objects confirmed to have large central cavities by submillimeter continuum emission. Consequently, dust filtration can explain the large-dust-depleted inner cavities seen in submillimeter observations, but cannot readily explain the small-dust-depleted inner cavities seen in near-infrared observations.

A few objects show different cavity sizes between near-infrared and submillimeter wavelengths. SAO 206462 shows a smaller cavity size at near-infrared wavelengths \citep[28 au;][]{gar13} than at submillimeter wavelengths \citep[39--46 au;][]{bro09,and11}. \citet{may12} also reported that RX J160421.7-213028 has a smaller cavity size at 1.6 $\mu$m polarized intensity (PI) images (63 au) than in the 880 $\mu$m continuum emission \citep[72 au;][]{mat12}.  \citet{gar13} proposed tidal interactions with planetary companions as an explanation for the observed diversity of cavity sizes. Different radial distributions of micron- and millimeter-size particles is thus key to investigating the dynamics of dust clearing in transitional disks. While modeling analysis based on SED is important, the spatial distribution of gas and various sizes dust in the disk must be evaluated by high-resolution direct imaging observations at multiple wavelengths to determine the physical mechanisms occurring at transitional phases. 

In this paper, we present the results of the first high-contrast near-infrared (1.6 $\mu$m) polarization imaging conducted on the transition disk associated with the young T Tauri star GM Aur by the Subaru 8.2 m Telescope at MaunaKea, Hawaii. GM Aur \citep[K5, 0.84$M_{\odot}$, 1--10 Myr old][]{sim00,har03} is located in the Taurus--Auriga molecular cloud, $\sim$140 pc away, initially revealed to have a rotating gaseous disk with an inner cavity \citep{koe93}. \citet{and11}, analyzing submillimeter (880 $\mu$m) interferometric images, derived the presence of a large inner cavity with a radius of 28 au. However, \citet{hor16}, examining visible wavelength ($\sim$0.54 $\mu$m) imaging data, reported a non-detection of the cavity. Our goal is to undertake a quantitative analysis of the spatial structure including the inner cavity to discuss the different aspects of the dust distribution than those already described in previous studies on the GM Aur disk. 

The High Contrast Instrument for the Subaru Next Generation Adaptive Optics \citep[HiCIAO;][]{tam06} provides high-resolution and high-contrast images, and has resolved more of the inner side of the disk than previous near-infrared HST/NICMOS observations. The results are different from the latest submillimeter observations regarding the cavity radius, and this morphological difference at different wavelengths may be interpreted as the result of physical interactions between the disk and unseen planets.

\section{OBSERVATIONS AND DATA REDUCTION}

Near-infrared (1.6$\mu$m) linear polarimetric differential images (PDI) of the GM Aur disk were obtained using the HiCIAO on the Subaru Telescope on the night of 2010 December 2. We used a double Wollaston prism to split the incident light into four 5$''\times$5\arcsec channels, two each of o- and e-ray sets (qPDI) to reduce the saturated radius under the expected inner cavity size ($r\sim0\farcs14$, 20 au at 140 pc). The imaging scale of HiCIAO in the qPDI mode is 9.5 mas per pixel. We obtained 18 data sets (four angular positions of the half-wave plate; 0\degr, 45\degr, 22\fdg5 and 67\fdg5 for one data set, to obtain full polarization coverage with minimal artifacts) with 8 s exposure for each frame. The total integration time was 576 s. This observation was carried out as part of the SEEDS \citep[Strategic Explorations of Exoplanets and Disks with Subaru;][]{tam09} project. 

The Image Reduction and Analysis Facility (IRAF)\footnote{The IRAF is distributed by the National Optical Astronomy Observatory, which is operated by the Association of Universities for Research in Astronomy under a cooperative agreement with the National Science Foundation.} with a custom script pipeline \citep{has11} was used for polarimetric data reduction in the standard manner of infrared image reduction. After bias subtraction and bad pixel correction, we obtained $+Q$, $-Q$, $+U$, and $-U$ images from e- and o-ray images at each angular position of the half-wave plate. The PI was given by $PI=\sqrt{Q^2+U^2}$.

Although a disk-like structure is visible in the PI image, the polarization vectors show a tendency to align toward the minor axis of the disk (Figure \ref{fig01}($a$)), not in circular symmetry, as expected from the Fresnel reflection. This is considered to be a residual stellar halo in the obtained images, because the standard reduction procedure cannot perfectly remove the convolved stellar point spread function. To obtain a more accurate disk-origin PI, we first computed the polarization halo by calculating the average polarization strength $P$  (1.27$\pm$0.05$\%$) and the average polarization angle $\theta$ (150\fdg5$\pm$0\fdg5) from the unsaturated Stokes {\it I}, {\it Q}, and {\it U} images. By subtracting the polarization halo model from the Stokes {\it Q} and {\it U} images, we obtained the halo-subtracted Stokes parameters, $Q_{\rm sub}$ and $U_{\rm sub}$. The final halo-subtracted polarized intensity ($\rm{PI}_{\rm sub}$) is computed via $\rm{PI}_{\rm sub}$=$\sqrt{Q^2_{\rm sub}+ U^2_{\rm sub}}$. Figure \ref{fig01}($b$) is the resultant $\rm{PI}_{\rm sub}$ image with polarization vectors, which exhibit circular symmetry. 

$H$-band angular differential images \citep[ADI;][]{mar06} were also obtained using the HiCIAO with 22 minutes of exposure time on 2011 December 22, to survey the presence of planets. The ADI data reduction was conducted with the LOCI algorithm \citep[Locally Optimized Combination of Images;][]{laf07}. In the LOCI process, each of the initial images is divided into concentric ring subsections, and the optimized background reference is then calculated for each subsection separately based on the counterpart subsection of the other images. The individual reference image is constructed from these locally optimized data for each individual initial image. This LOCI data reduction was conducted as part of a comprehensive study on the SEEDS high-contrast imaging survey of exoplanets, and detailed parameters are described in \citet{uya16}.

\section{RESULTS}

The final {\it H}-band PI image of the GM Aur disk with a software mask ($r\sim$40 mas) is shown in Figure \ref{fig01}($b$). The symmetric and elliptical disk structure is resolved. We fitted elliptical isophotes on the resultant images to measure the inclinations and position angles (PAs) of the disk. The elliptical fitting results are shown in Table \ref{gma_tab}, and overlaid on the image.  We found a significant offset between the disk center and the location of the central star of $\sim$38 mas at a PA of 252\degr, which is roughly along the minor axis. This offset indicates that the southeast side is inclined toward us \citep[e.g.,][]{don12b}.

Figure \ref{fig02} shows contour map image and radial PI profiles along the major and minor axes. We fitted a power law to each slope and found a clear evidence of an inner cavity with a radius of 18$\pm2$ au. This is the first detection of a cavity in the GM Aur disk at near-infrared wavelengths, and the cavity radius is smaller than the results of observations at submillimeter wavelengths of 28 au \citep{and11}. However, previous detailed modeling by \citet{esp10} employed an inner cavity of $\sim$20 au that is in good agreement with our result.

The results of ADI/LOCI analysis on GM Aur are shown in Figure \ref{fig03}. No significant point source is seen in the ADI/LOCI image (Figure \ref{fig03}($a$)). The signal-to-noise map (Figure \ref{fig03}($b$)) shows that black and white patterns around the mask in the resultant image are just speckle noises. Figure \ref{fig03}($c$) shows the companion mass limit that we could detect at 5$\sigma$ from the resultant image. The mass was converted from the contrast by assuming the COND evolutionary model \citep{bar03}. We also took into account the self-subtraction effect of the LOCI algorithm \citep[See Section 3 in ][]{uya16}. The detectable mass limit is only a few {\it M}$_{\rm Jup}$; therefore, the presence of companions with the masses of typical brown dwarfs is excluded around GM Aur.

\section{DISCUSSION}

\subsection{Different Brightness Slope: Polarization and Non-polarization}

The PI profiles of the outer disk along the two major axes (Figure \ref{fig02}) are well fitted to a power law with power indices of $-$1.62$\pm$0.04 and $-$1.82$\pm$0.06 for the northeast and southwest axes, respectively, which indicates a flared disk surface \citep[a flat disk is expected to have a steeper profile, e.g.,][]{whi92}. We found that the power indices of the minor axes, $\leq$1.6, are not consistent with those of the radial brightness intensity ($I$) profiles obtained from HST/NICMOS observations \citep{sch03}, namely a power index of $-$3.5 for a F160W band filter. Since the wavelength difference between the HiCIAO {\it H}-band filter (1.3--1.9 $\mu$m) and the NICMOS F160W band filter (1.4--1.8 $\mu$m) is negligible for surface brightness comparison, the power index difference between PI and $I$ profiles indicate a change in polarization efficiency at different radii. The polarization efficiency is determined by many variables such as the composition of particle shapes and sizes, multiple scattering, and changes in scattering angle by flared surfaces  \citep[e.g.,][]{whi02,per09}. Some of the physical parameters mentioned above increase the polarization efficiency as the radius increases. Therefore, the power index difference between PI and $I$ profiles may be a sign of various grain growth process at different radii. To obtain more robust explanation for polarization efficiency variance with radius, multi-band observations \citep[e.g.,][]{per13} and detailed model comparison \citep[e.g.,][]{min12} must be conducted.

\subsection{Imaging Diagnostics: Cavity Edge Radius and Planet Mass}\label{juan}

Recent imaging observations of disks have revealed spatial differentiation in the grain sizes in transitional disks, since observations at different wavelengths trace different grain sizes, showing different spatial structures of the disks \citep[a missing cavity at near-infrared imaging;][]{don12a}. \citet{zhu12} and \citet{pin12} suggested that the dust filtration effect by planet-induced cavity edges may cause dust particles at different radii to have different grain sizes. \citet{jua13} performed 2-dimensional hydrodynamical and dust evolution models combined with instrument simulations of VLT/SPHERE-ZIMPOL (0.65 $\mu$m), Subaru/HiCIAO (1.6 $\mu$m), and ALMA (850 $\mu$m). Their results provide a simple mass estimating function using the ratio of SPHERE ZIMPOL $R$-band cavity edge radius to ALMA Band 7 peak radius, and it allows predictions to be made for emitted/scattered light spatial images at different wavelengths for several cases of planet masses and locations.

To estimate possible planet masses in the GM Aur disk with cavity sizes at near-infrared (this work) and submillimeter wavelengths \citep[28 au;][]{and11}, we assumed that (1) an 18 au radius cavity seen in a HiCIAO infrared polarized intensity image is induced by a planet at 20 au separation from the central star, that (2) the cavity edge radius seen in ZIMPOL can be replaced with that seen in HiCIAO, since ZIMPOL and HiCIAO polarized intensity images trace similar dust particle sizes (1--10$\mu$m in general) \citep{jua13}, and (3) HiCIAIO can resolve a thin gap and an inner disk with the radius of 20 au or less \citep[e.g.,][]{aki15}. Therefore, it should be noted that this is only a rough estimation. Consequently, we obtained 3--4 $M_{\rm Jup}$ as the mass of a possible planet at 20 au separation. The possible planet could not be detected in the ADI/LOCI resultant image due to a larger inner working angle and higher detectable mass limit near the central star (Figure \ref{fig03}($c$)). Even if we had obtained sufficient contrast to detect the planet, the optically thick inner edge of the disk would have concealed the planet under the dust. Therefore, to reveal the planet inside the dust, an additional method may be necessary, such as direct detection of H$\alpha$ emission from the gas accreting onto the protoplanet \citep[e.g., LkCa 15;][]{sal15}.

\subsection{Origin of the Inner cavity}

To explain the near-infrared deficit in SED, which is indicative of an inner cavity in the disk, a number of physical mechanisms have been suggested, such as photoevaporation, grain growth, disk wind by magnetorotational instability, and gravitational interactions between disks and planets. However, to date, none of the suggested mechanisms have clearly explained the formation of an infrared inner cavity in transitional disks, including the GM Aur disk. 

Photoevaporation is a dominant mechanism when the mass accretion rate is sufficiently low. However, the high mass accretion rate of the GM Aur disk \citep[4$\times$10$^{\rm -9}$-1$\times$10$^{\rm -8}$$M_{\odot}$;][]{esp07a,ing15} requires a substantial supply of material from a massive outer disk \citep[$\sim$0.16$M_{\odot}$;][]{hug09}. {Grain growth \citep[e.g.,][]{bir09} should proceed without reducing the gas density, but \cite{dut08} revealed the presence of a large gas cavity in the GM Aur disk from a lack of CO line emission. Furthermore, without an additional process, it cannot explain the depletion of small dust in transitional disks due to continuous replenishing of small dust via fragmentation \citep{bir09}. Inside-out evacuation via disk wind \citep{chi07} cannot be reconciled with the silicate emission at 10$\mu$m in the IRS spectrum of GM Aur \citep{cal05}, which indicates the existence of $\sim$0.02 lunar masses of $\mu$m-size dust in the inner region within $\sim$ 5 au \citep{cal05}. We note that this faint inner disk is outside of our scope. \cite{ros14} proposed that radial flow of gas with near free-fall velocity can explain the dust depletion at inner cavities that survive even for a relatively high mass accretion rate from the outer disk (fast radial flows model). However, this scenario has not resolved how gas in the disk cavity efficiently shed its angular momentum. A Puffed-up inner disk can produce a cavity-like shape by self-shadowing \citep[e.g.,][]{gar14}, but it is not consistent with the optically thin inner disk of GM Aur.

A dust filtration effect \citep{ric06} by a planet-induced gap is the most likely mechanism to explain the inner cavity in the GM Aur transitional disk \citep{hug09}. Recently, \cite{zhu12} found that the gap outer edge near a 6 $M_{\rm{Jup}}$ planet can only filter particles that are 0.1 mm and larger, and proposed that the dust filtration+grain growth scenario can explain the strong near-infrared deficit in SED of the GM Aur  disk. However, those scenarios commonly have difficulties explaining the lack of micron-size particles in the central region, which we have reported in this work. Therefore, although dust filtration and grain growth may play key roles in inner disk clearing, there must be an additional \deleted{process, such as interactions with} factor, such as the presence of multiple planets \citep{don15}.

\section{CONCLUSION}

We have presented a spatially resolved image of the GM Aur transitional disk at near-infrared wavelengths, obtained using Subaru/HiCIAO, and have revealed a large inner cavity, which is seen for the first time at these wavelengths. The measured cavity radius of 18 au is significantly smaller than that of the latest measurement at submillimeter wavelengths of 28 au \citep{and11}. This discrepancy in cavity size may be caused by a planet embedded in the cavity, the mass of which has been estimated at 3--4 $M_{\rm Jup}$ with some assumptions. We also suggested various grain growth process at different radii based on the difference between PI and $I$ radial profiles.

The physical mechanisms of a large inner cavity surrounded by a massive outer disk are still not well understood. In particular, the mechanism for clearing a near-infrared cavity in a transitional disk has not yet been determined. For a submillimeter inner cavity, a dust filtration mechanism is the most likely hypothesis. However, our discovery of a near-infrared inner cavity suggests that the dust filtration effect should be combined with an additional disk clearing processes, or that micron-size particles are not coupled well with the accreting gas for some reason. High-resolution imaging observations by ALMA will be important for verifying the dust filtration scenario by tracing the gas flow in the near-infrared cavity in the GM Aur disk.

\

We are grateful to the referee for providing useful comments that led to an improved version of this letter. M.T. and J.C. are supported by a Grant-in-Aid for Scientific Research (No.15H02063) and by the U.S. National Science Foundation under award{\footnotesize\bf\#}1009203, respectively.

\begin{deluxetable}{ccc}
\tablecaption{Geometric measurements of GM Aur disk \label{gma_tab}}
\tablehead{
\colhead{Parameter} & \colhead{Outer Edge\tablenotemark{a}} & \colhead{Inner Edge\tablenotemark{b}}
}
\startdata
Radius of semi-major axis (au) & $70\pm4$ & $18\pm2$\\
Position angle of semi-major axis (\degr)\tablenotemark{c} & $59\pm2$ & $53\pm2$ \\
Inclination (\degr)\tablenotemark{d} & $64\pm2$ & - \\
Geometric center offset (mas)\tablenotemark{e} & ($-12\pm5$,$-36\pm6$) & ($-8\pm1$,$-4\pm1$) \\
\enddata
\tablenotetext{a}{The positions where the flux becomes noise level were obtained first from radial profiles at position angles every 10\degr. Then least-squares estimation elliptical fitting with five free geometric parameters was conducted using the IRAF package stsdas.analysis.}
\tablenotetext{b}{Based on the flux peaks in the radial profiles. The elliptical fitting was not available at the inner edge because the number of peak points was insufficient to fit with a small error.}
\tablenotetext{c}{Counterclockwise from the north axis.}
\tablenotetext{d}{Derived from the ellipticity. The inclination of a face-on disk is 0\degr.}
\tablenotetext{e}{($\Delta$R.A., $\Delta$decl.). The origin of the coordinates corresponds to the position of the central star.}
\end{deluxetable}

\begin{figure}[h]
\figurenum{1}
\plotone{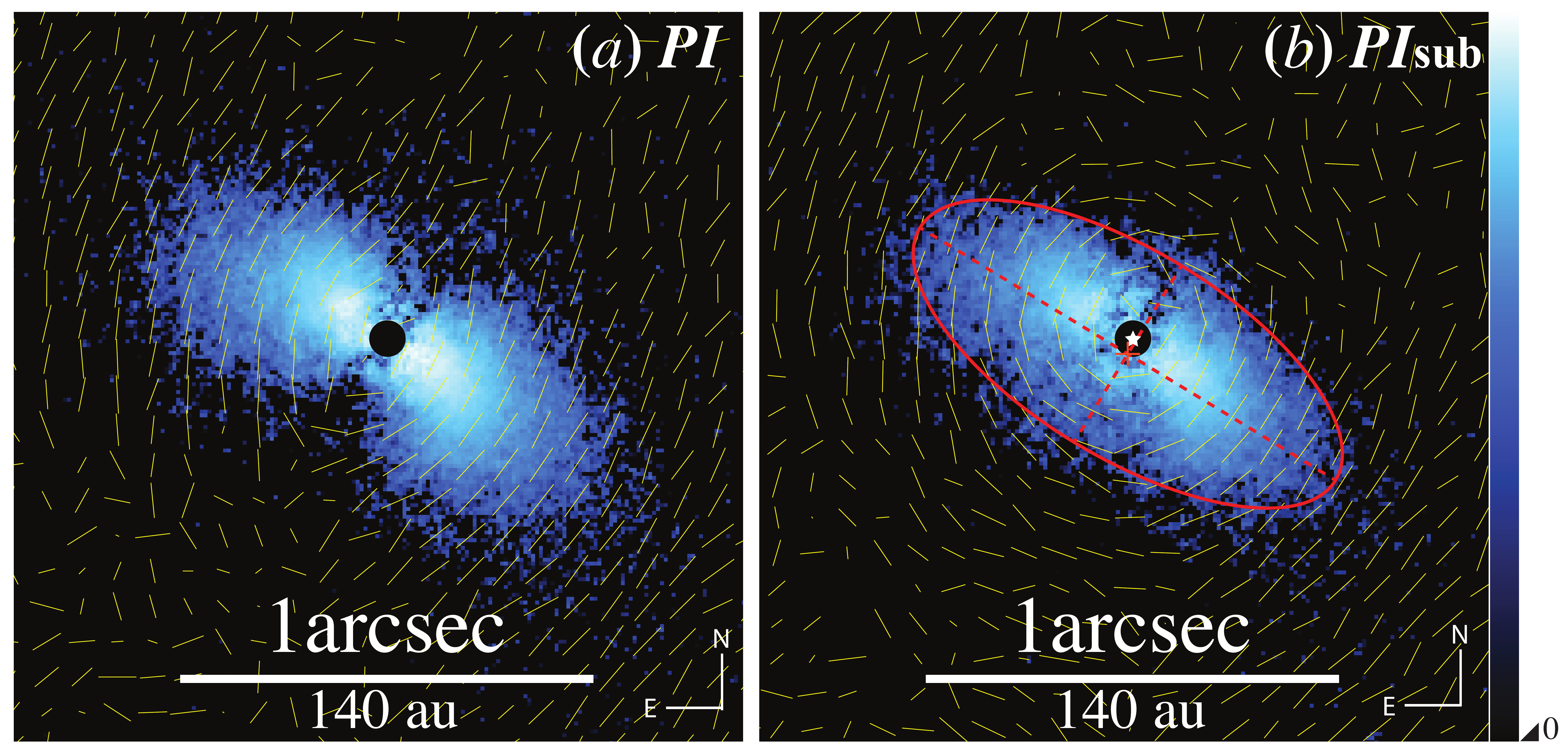}
\caption{PI and overlapped polarization vector map images (a) before and (b) after halo subtraction. The triangle on the color bar indicates the zero level in the color scale. The saturated region is occulted by a software mask ($r = 5$ pixels $\sim$ 0\farcs05), the vectors are binned with spatial resolution, and the lengths are arbitrary for presentation purposes. The red markings and the white star represent the elliptical fitting results and the location of the star, respectively. (a) The effect of the polarization halo appears to tend toward the minor axis of the disk. (b) The polarization tendency toward the minor axis is removed, and a disk-origin polarization along the disk surface is revealed.}\label{fig01}
\end{figure}

\begin{figure}[h]
\figurenum{2}
\plotone{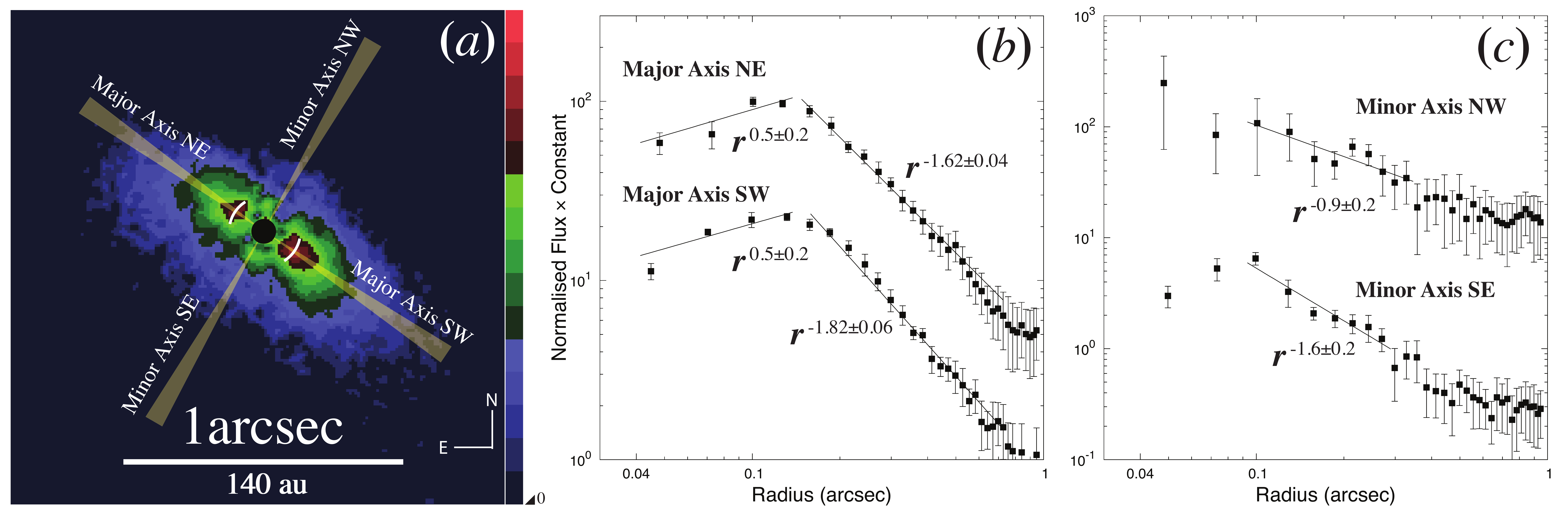}
\caption{(a) Contour map image. The inner edge of the disk (white lines) and the central inner cavity (between the red peaks and software mask) are more clear in this image. The image has been smoothed by a Gaussian with $r = 2$ pixels. (b) and (c) Radial profiles with 1$\sigma$ error bars at yellow hatched regions of the minor and major axes in (a). The values were measured at each axis within the range $\pm$5\degr, and were binned with a width of $dr = 4$ pixels.}\label{fig02}
\end{figure}

\begin{figure}[h]

\figurenum{3}
\plotone{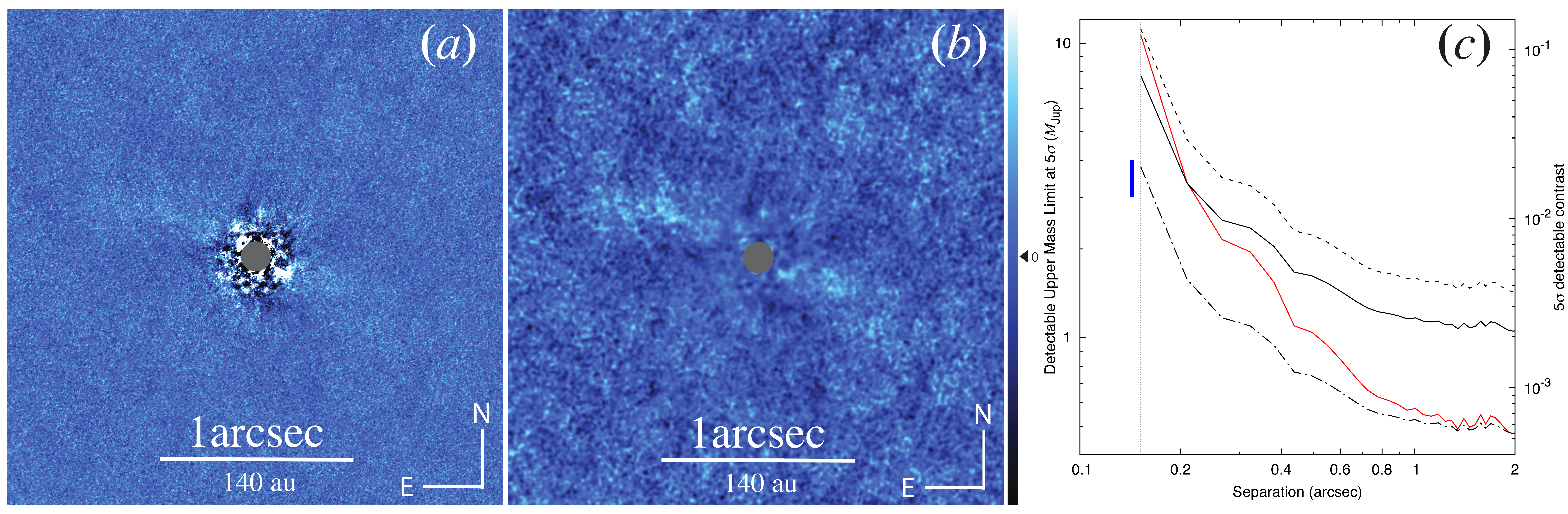}
\caption{(a) The $H$-band ADI/LOCI resultant image. (b) Signal-to-noise map at a stretch of [-5,5]$\sigma$. (c) Detectable upper contrast (red) and mass (black) limit at 5$\sigma$ based on the ADI/LOCI resultant image, assuming ages of 1, 5, and 10 Myr (dashed--dotted, solid, and dashed curves, respectively). The vertical dotted line is the mask radius. The blue thick line represents the assumed planet (3--4 {\it M}$_{\rm Jup}$ at 20 au separation; see Section \ref{juan}).}\label{fig03}
\end{figure}

\clearpage

\allauthors

\listofchanges

\end{document}